\documentclass[11pt]{article}
\usepackage{moriond,epsfig}

\def\be{\begin{equation}}
\def\ee{\end{equation}}
\def\bea{\begin{eqnarray}}
\def\eea{\end{eqnarray}}

\begin{document}
\vspace*{4cm}
\title{\boldmath$b\to s$ transitions: where to look for SUSY
\footnote{Talk given by M.C.}}

\author{M. Ciuchini$^1$, E. Franco$^2$, A. Masiero$^3$, and L. Silvestrini$^2$}

\address{$^1$INFN Sezione di Roma III and Dip. di Fisica,\\
Univ. di Roma Tre, Via della Vasca Navale 84, I-00146 Rome, Italy.\\
$^2$INFN Sezione di Roma and Dip. di Fisica, Univ. di Roma ``La Sapienza'',\\
P.le A. Moro 2, I-00185 Rome, Italy.\\
$^3$Dip. di Fisica ``G. Galilei'', Univ. di Padova and INFN,\\
Sezione di Padova, Via Marzolo 8, I-35121 Padua, Italy.
}

\maketitle\abstracts{We present a systematic
  study of the CP conserving and violating SUSY contributions to $b
  \to s$ processes in a generic MSSM, considering gluino exchange
  contributions. 
  Experimental information on $B \to X_s 
  \gamma$, $B\to X_s \ell^+ \ell^-$ and the $B_s - \bar 
  B_s$ mass difference $\Delta M_s$ have been taken into account. 
  We study the induced correlations among several
  observables: $\Delta M_s$ and the amount of CP violation in $B \to \phi
  K_s$, $B_s \to J/\psi \phi$, $B \to X_s \gamma$. Our results show that     
  $b \to s$ transitions represent a splendid opportunity to
  constrain different MSSM realizations and
  offer concrete prospects to exhibit SUSY signals at $B$
  factories and hadron colliders.}

\section{Introduction}
Flavour physics is a very stringent test of
SUSY extensions of the Standard Model: in its general form, the Minimal Supersymmetric Standard Model (MSSM) can cause Flavour Changing Neutral Current and CP violating processes to arise at a rate
much higher than what is experimentally
observed~\cite{antichibagger}.

A closer look at the Unitarity Triangle
fit reveals that new physics (NP)
contributions to $s \to d$ and $b \to d$ transitions are strongly
constrained, while new contibutions to $b \to s$ transitions
affect the fit only if they interfere destructively with the SM
amplitude for $B_s - \bar B_s$ mixing, bringing the mass difference
below the present lower bound.  Other
processes not involved in the UT fit, for example the celebrated
$B \to X_s\gamma$, can provide constraints on any NP in $b \to s$ 
transitions. However, $B \to X_s \gamma$ mostly constrains the helicity
flipping contributions to the $b \to s$ transition. As we shall
see in the following, plenty of room is left for SUSY contributions to
interesting observables in this sector.

In these proceedings, we summarise the results of the analysis presented in
ref.~\cite{Ciuchini:2002uv}. We refer the interested reader to the
original paper for all the details omitted here.
This analysis aims at studying systematically
SUSY contributions to the (CP conserving and violating) $b \to s$
transitions in the context of a {\it generic} MSSM model with R
parity at a level of accuracy comparable to the SM UT fit (i.e.~using NLO QCD corrections and Lattice QCD hadronic matrix elements wherever possible).

We keep our analysis in the MSSM as general as possible. 
Minimality refers here only
to the minimal amount of superfields needed to supersymmetrize the SM
and to the presence of R parity. Otherwise the soft breaking terms are
left completely free and constrained only by phenomenology.
Technically the best way we have to account for the SUSY FCNC
contributions in such a general framework is via the mass insertion
method using the leading gluino exchange contributions~\cite{hall}.
In the Super-CKM basis, SUSY FCNC and CP violation arise from
off-diagonal terms in squark mass matrices only. These are conveniently
expressed as $(\delta_{ij})_{AB}\equiv (\Delta_{ij})_{AB}/m^2_{\tilde
  q}$, where $(\Delta_{ij})_{AB}$ is the mass term connecting squarks
of flavour $i$ and $j$ and ``helicities'' $A$ and $B$, and
$m_{\tilde q}$ is the average squark mass. In the absence of any
horizontal symmetry and for a generic SUSY breaking mechanism, one
expects $(\delta^d_{ij})_{LL} \leq \mathcal{O} (1)$,
$(\delta^d_{ij})_{RR} \leq \mathcal{O} (1)$, $(\delta^d_{ij})_{LR}
\leq \mathcal{O} (m_{d_k}/m_{\tilde q})$ and $(\delta^d_{ij})_{RL}
\leq \mathcal{O} (m_{d_k}/m_{\tilde q})$, with $k=$max$(i,j)$. The
last two inequalities are also imposed by the requirement of avoiding
charge and colour breaking minima as well as unbounded from below
directions in scalar potentials~\cite{Casas:1996de}.

Detailed analyses carried out in SUSY have shown that one must have $(\delta^d_{12})_{AB}$ and $(\delta^d_{13})_{AB}$ much smaller than what 
naively
expected~\cite{Becirevic:2001jj,Ciuchini:1998ix}. It is therefore
reasonable to assume that
$(\delta^d_{12})_{AB}\sim (\delta^d_{13})_{AB}\sim 0$. Under this assumption, we present constraints on $(\delta^d_{23})_{AB}$
from available data and possible effects in present and future
measurements.

\section{Phenomenological Analysis}
Our analysis aims at determining the allowed
regions in the SUSY parameter space governing $b \to s$ transitions,
studying the correlations among different observables and pointing out
possible signals of SUSY. The constraints on the parameter space come
from:

1. The BR$(B \to X_s \gamma)=(3.29 \pm 0.34)\times
  10^{-4}$ (experimental results as reported in~\cite{Stocchi:2002yi},
  rescaled according to ref.~\cite{Gambino:2001ew}). 
  
2. The CP asymmetry $A_{CP}(B \to X_s \gamma)=-0.02 \pm
  0.04$~\cite{Stocchi:2002yi}.

3. The BR$(B \to X_s \ell^+ \ell^-)=(6.1 \pm 1.4 \pm 1.3)\times
  10^{-6}$~\cite{Stocchi:2002yi}.

4. The lower bound on the $B_s - \bar B_s$ mass difference $\Delta
  M_{B_s} > 14.4$ ps$^{-1}$ \cite{Stocchi:2002yi}.

We have also considered BR's and CP asymmetries for $B \to K \pi$ and
found that, given the large theoretical uncertainties, they give no significant constraints on the $\delta$'s.

For $B \to \phi K_s$, we have studied the BR and the
coefficients $C_{\phi K}$ and $S_{\phi K}$ of cosine and sine terms in
the time-dependent CP asymmetry.

All the details concerning the treatment of the different amplitudes
entering the analysis can be found in ref.~\cite{Ciuchini:2002uv}. In summary, we use:

{\it i) $\Delta B=2$ amplitudes.} Full NLO SM and LO gluino-mediated matching condition, NLO QCD evolution and hadronic matrix elements from
lattice calculations.

{\it ii) $\Delta B=1$ amplitudes.} Full NLO SM and LO gluino-mediated matching condition and NLO QCD evolution. The matrix elements of
semileptonic and radiative decays include $\alpha_s$ terms, Sudakov 
resummation, and the first corrections suppressed by powers of the heavy quark masses. For non-leptonic decays, such as $B \to K \pi$ and $B \to \phi K_s$, we adopt BBNS factorization~\cite{BBNS}, with  an enlarged
range for the annihilation parameter $\rho_A$, in the
spirit of the criticism of ref.~\cite{Charming}. This choice
maximizes the sensitivity of the factorized amplitudes to SUSY contributions, which is expected to be much lower if the
power corrections are dominated by the ``charming penguin'' contributions~\cite{Ciuchini:1997hb}.

Another source of potentially large SUSY effects in
$B \to \phi K_s$ is the contribution of the chromomagnetic
operator which can be
substantially enhanced by SUSY without spoiling the experimental
constraints from $B\to X_s\gamma$~\cite{c8g}. 
Indeed, the time-dependent asymmetry in $B \to \phi K_s$
is more sensitive to the SUSY parameters in the case of
chirality-flipping insertions which
enter the amplitude in the coefficient of the chromomagnetic operator.
One should keep in mind, however, that the corresponding
matrix element, being of order $\alpha_s$, has large uncertainties
in QCD factorization.

We performed a MonteCarlo analysis, generating weighted random
configurations of input parameters (see ref.~\cite{Ciuchini:2000de}
for details of this procedure) and computing for each configuration
the processes listed above. We study the clustering induced by the
contraints on various observables and parameters, assuming that each
unconstrained $\delta_{23}^d$ fills uniformly a square $(-1\dots 1$,
$-1\dots 1)$ in the complex plane. The ranges of CKM parameters have
been taken from the UT fit ($\bar \rho=0.178 \pm
0.046$, $\bar \eta=0.341 \pm 0.028$), and hadronic parameter ranges
are those used in ref.~\cite{Ciuchini:2002uv}.

Concerning SUSY parameters, we fix $m_{\tilde q}=m_{\tilde g}=350$ GeV
and consider different possibilities for the mass insertions. In
addition to studying single insertions, we also examine the
effects of the left-right symmetric case
$(\delta^d_{23})_{LL}=(\delta^d_{23})_{RR}$. 

\begin{figure}[t]
  \begin{center}
    \begin{tabular}{c c}
      \includegraphics[width=0.4\textwidth]{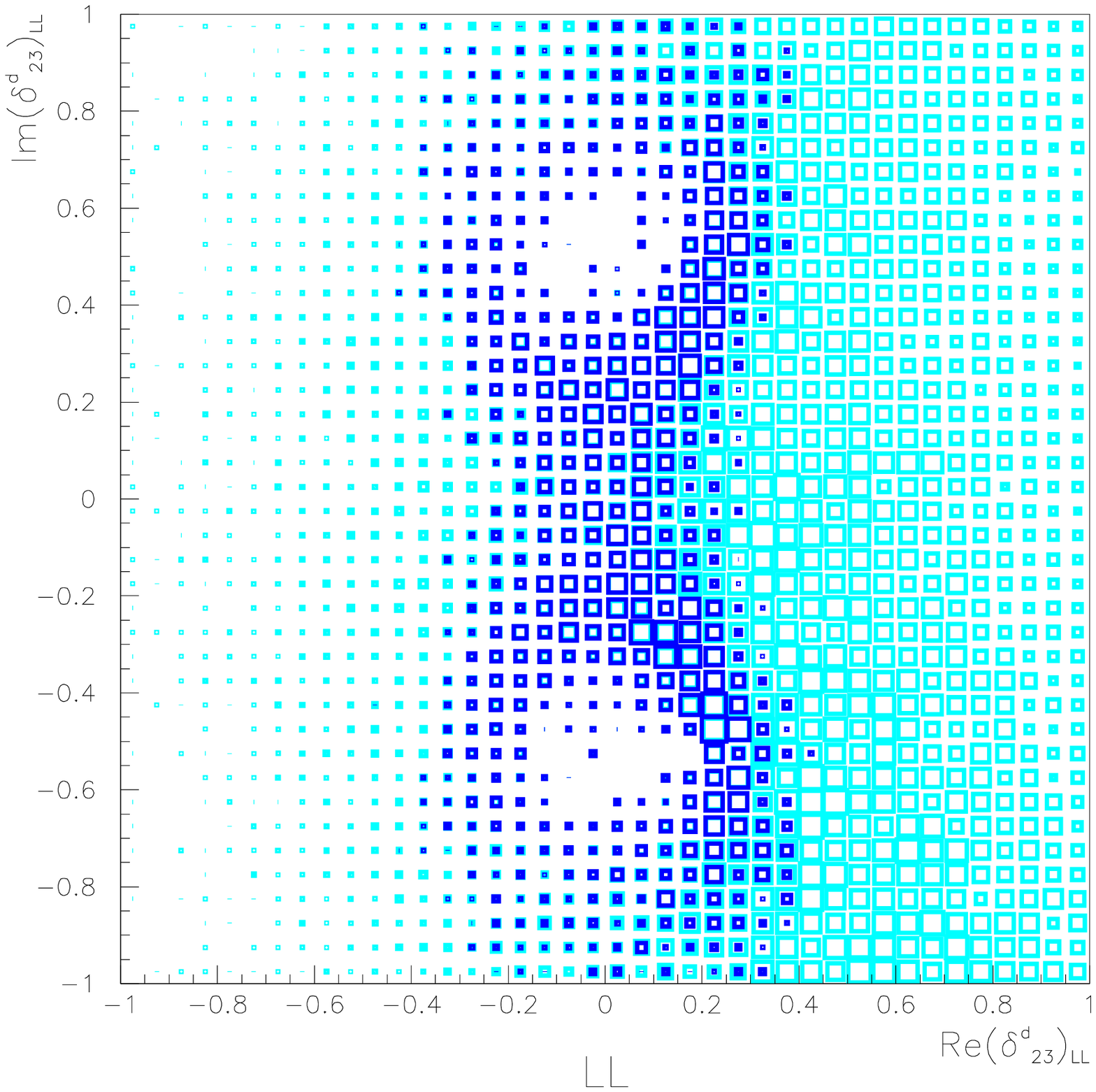} &
      \includegraphics[width=0.4\textwidth]{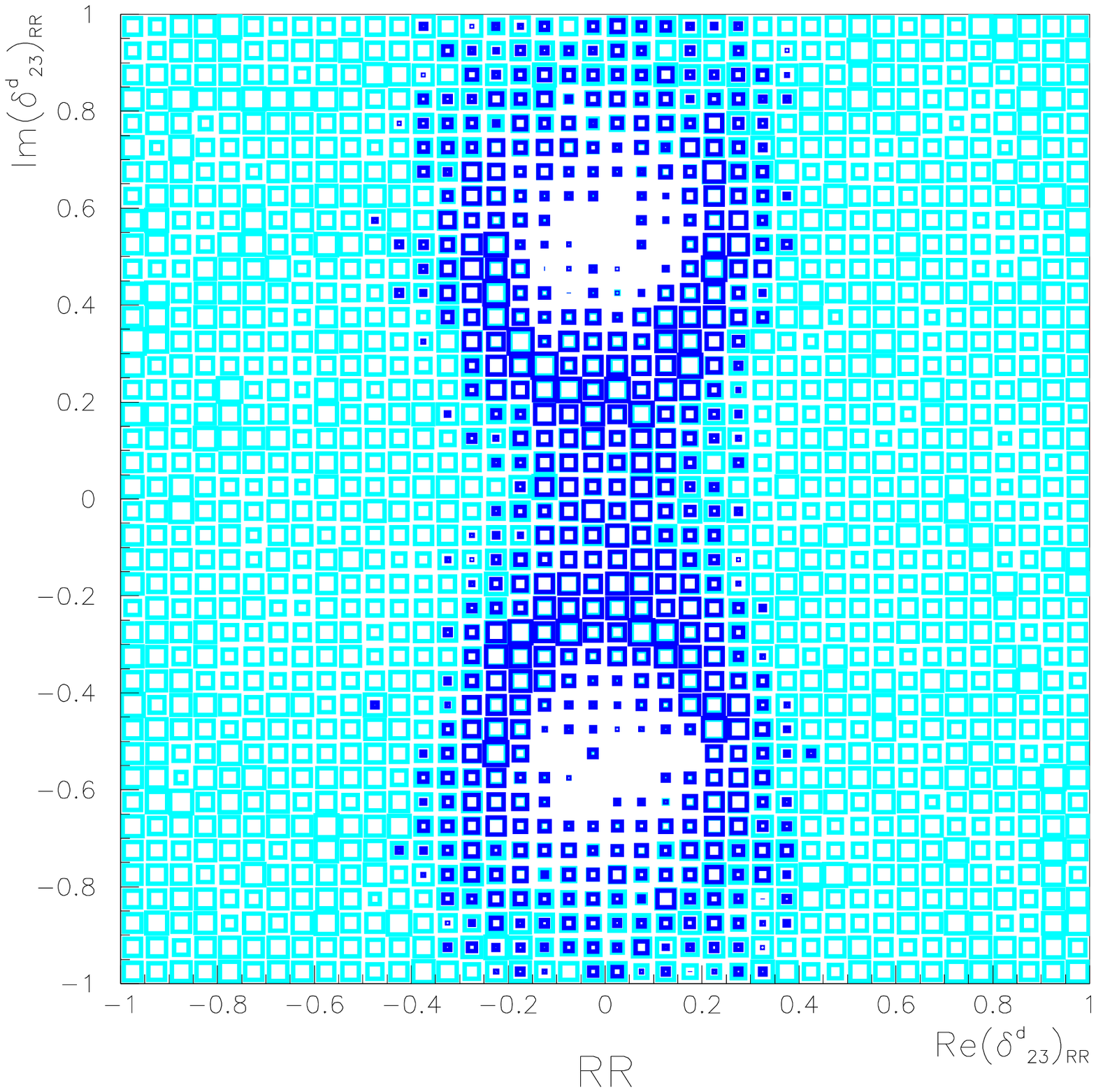} \\
      \includegraphics[width=0.4\textwidth]{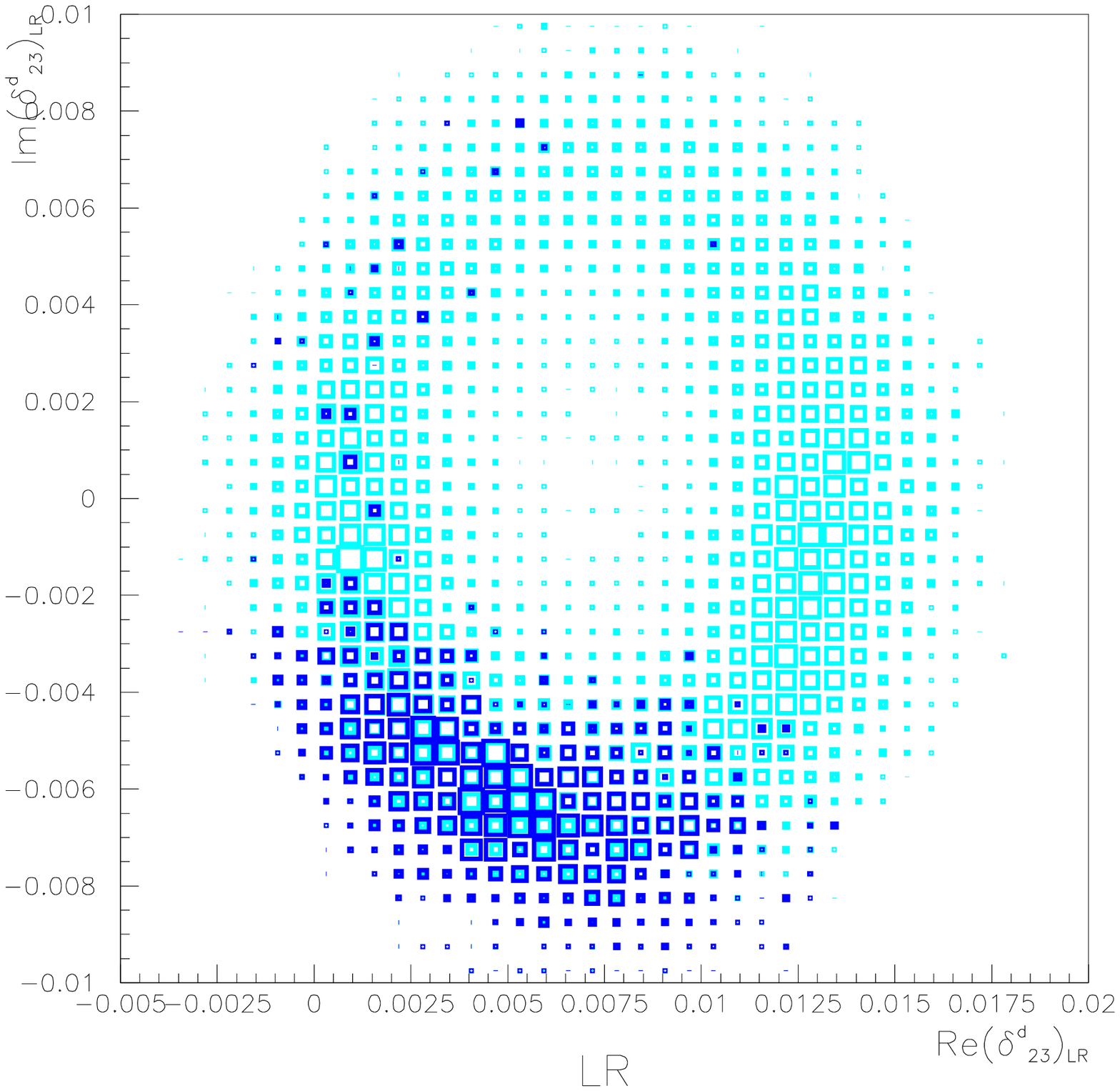} &
      \includegraphics[width=0.4\textwidth]{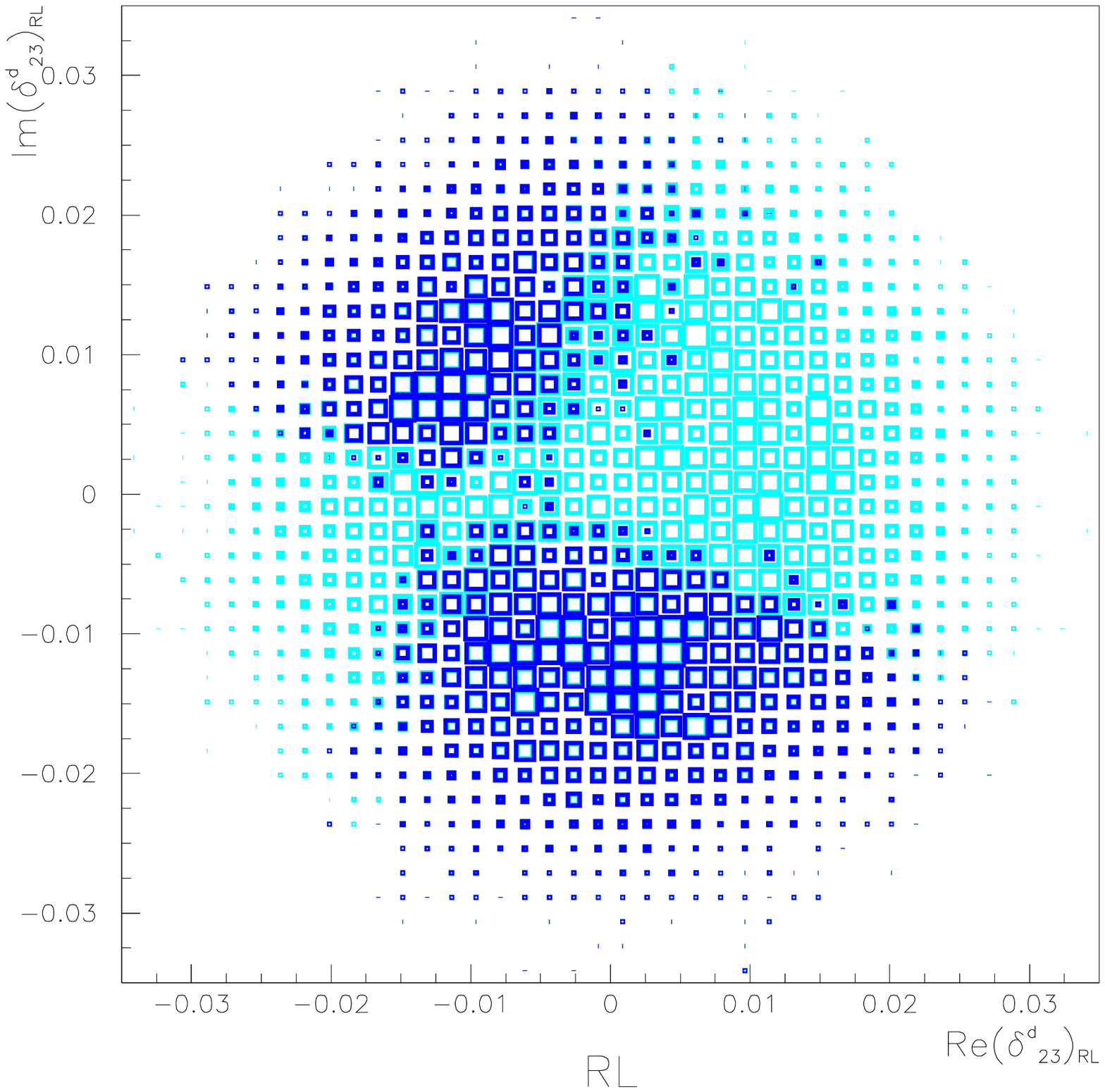} \\ 
    \end{tabular}
  \end{center}
  \caption{Allowed regions in the  
    Re$(\delta^d_{23})_{AB}$--Im$(\delta^d_{23})_{AB}$ space for
   $AB=(LL,RR,LR,RL)$. 
    The darker
    regions are selected imposing $\Delta
    m_s<20$ ps$^{-1}$ for
    $LL$ and $RR$ insertions and $S_{\phi K}<0$ for
    $LR$ and $RL$ insertions.}
  \label{fig:ranges1}
\end{figure}

In fig.~\ref{fig:ranges1} we display the clustering of events in the
Re$(\delta^d_{23})_{AB}$--Im$(\delta^d_{23})_{AB}$ plane in the single
insertion case. Here and in the following plots, larger boxes
correspond to larger numbers of weighted events. Constraints from
$BR(B\to X_s\gamma)$, $A_{CP}(B\to X_s\gamma)$, $BR(B\to X_s \ell^+
\ell^-)$ and the lower bound on $\Delta M_s$ have been applied, as
discussed above. The darker regions are selected imposing the further
constraint $\Delta M_s<20$~ps$^{-1}$ for $LL$ and $RR$ insertions and
$S_{\phi K}<0$ for $LR$ and $RL$ insertions.  For helicity conserving
insertions, the constraints are of order $1$. A significant reduction
of the allowed region appears if the cut on $\Delta M_s$ is imposed.
The asymmetry of the $LL$ plot is due to the interference with the SM
contribution. In the helicity flipping cases, constraints are of order
$10^{-2}$. For these values of the parameters,
$\Delta M_s$ is unaffected. We show the effect of requiring $S_{\phi
  K}<0$: it is apparent that a nonvanishing Im$\,\delta_{23}^d$ is
needed to meet this condition.

\begin{figure}[t]
  \begin{center}
    \begin{tabular}{c c}
      \includegraphics[width=0.4\textwidth]{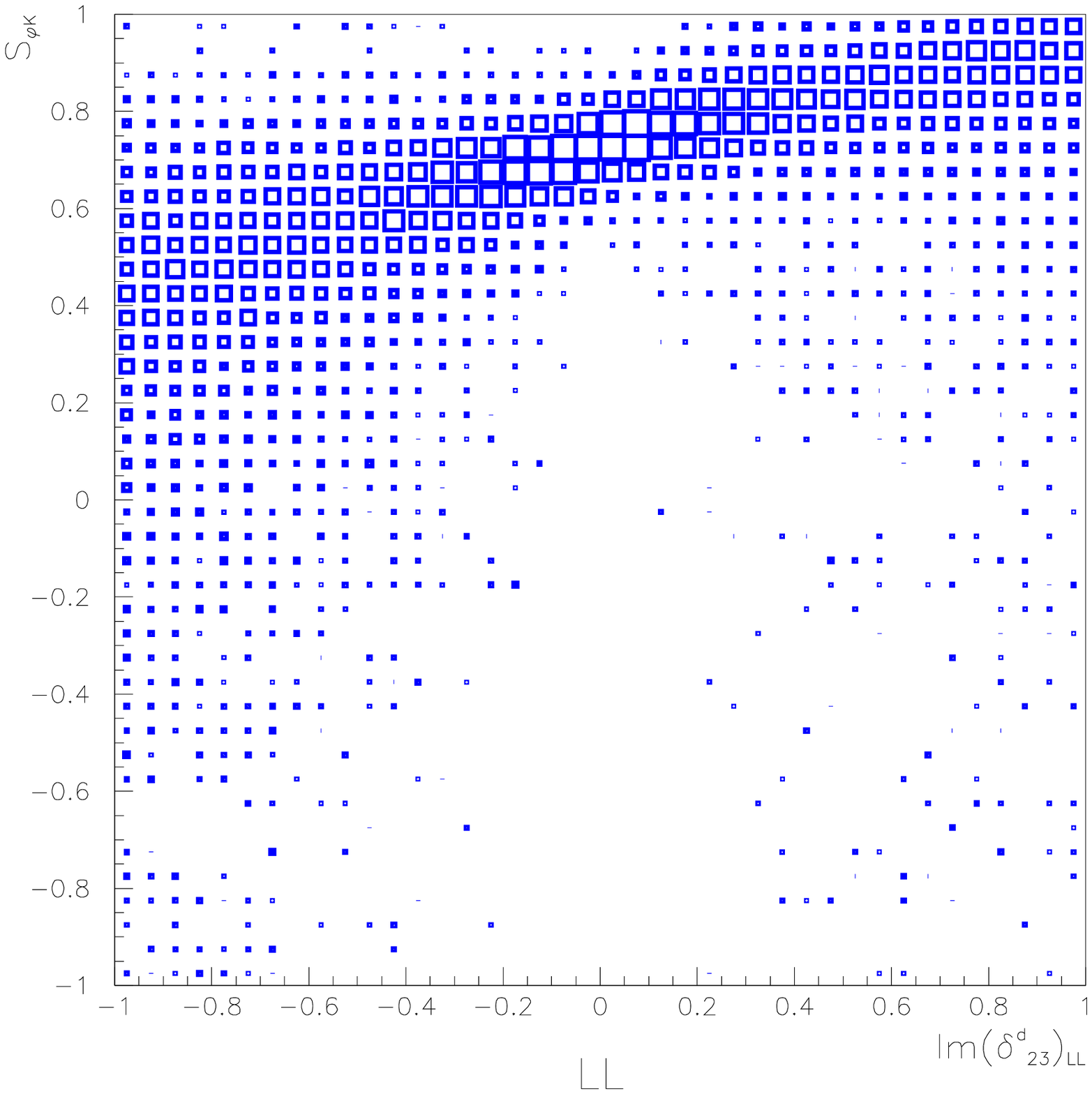} &
      \includegraphics[width=0.4\textwidth]{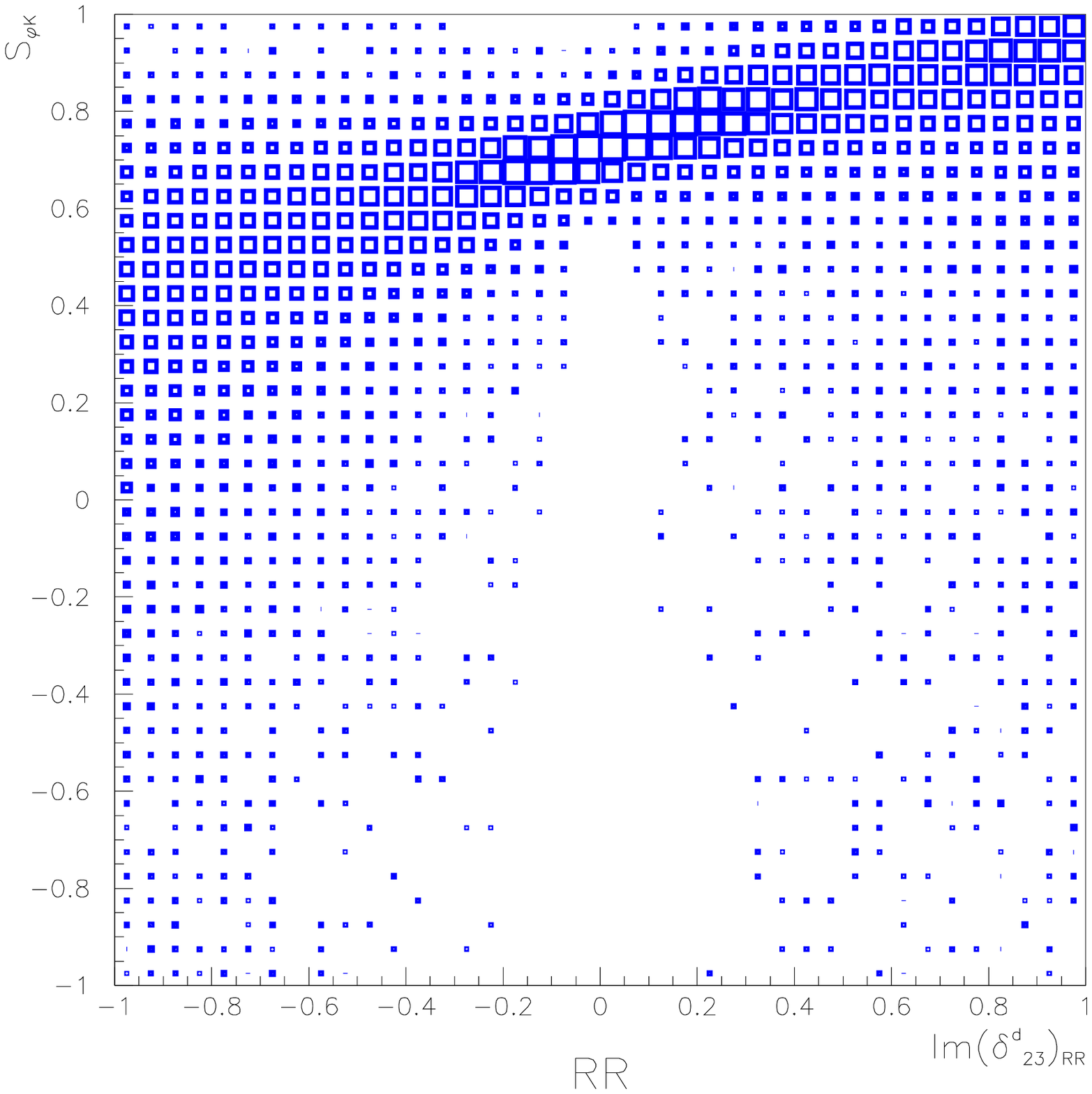} \\
      \includegraphics[width=0.4\textwidth]{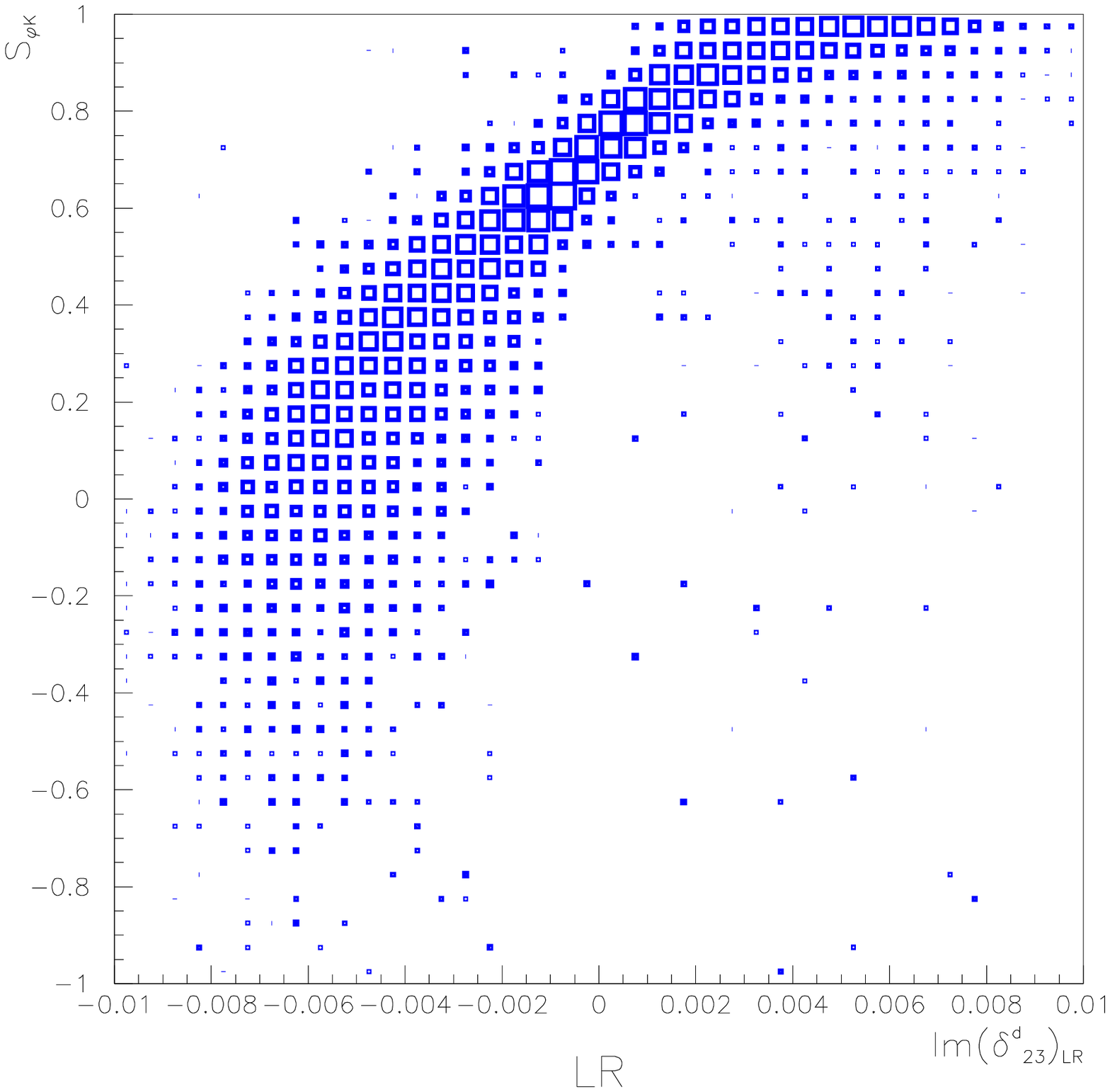} &
      \includegraphics[width=0.4\textwidth]{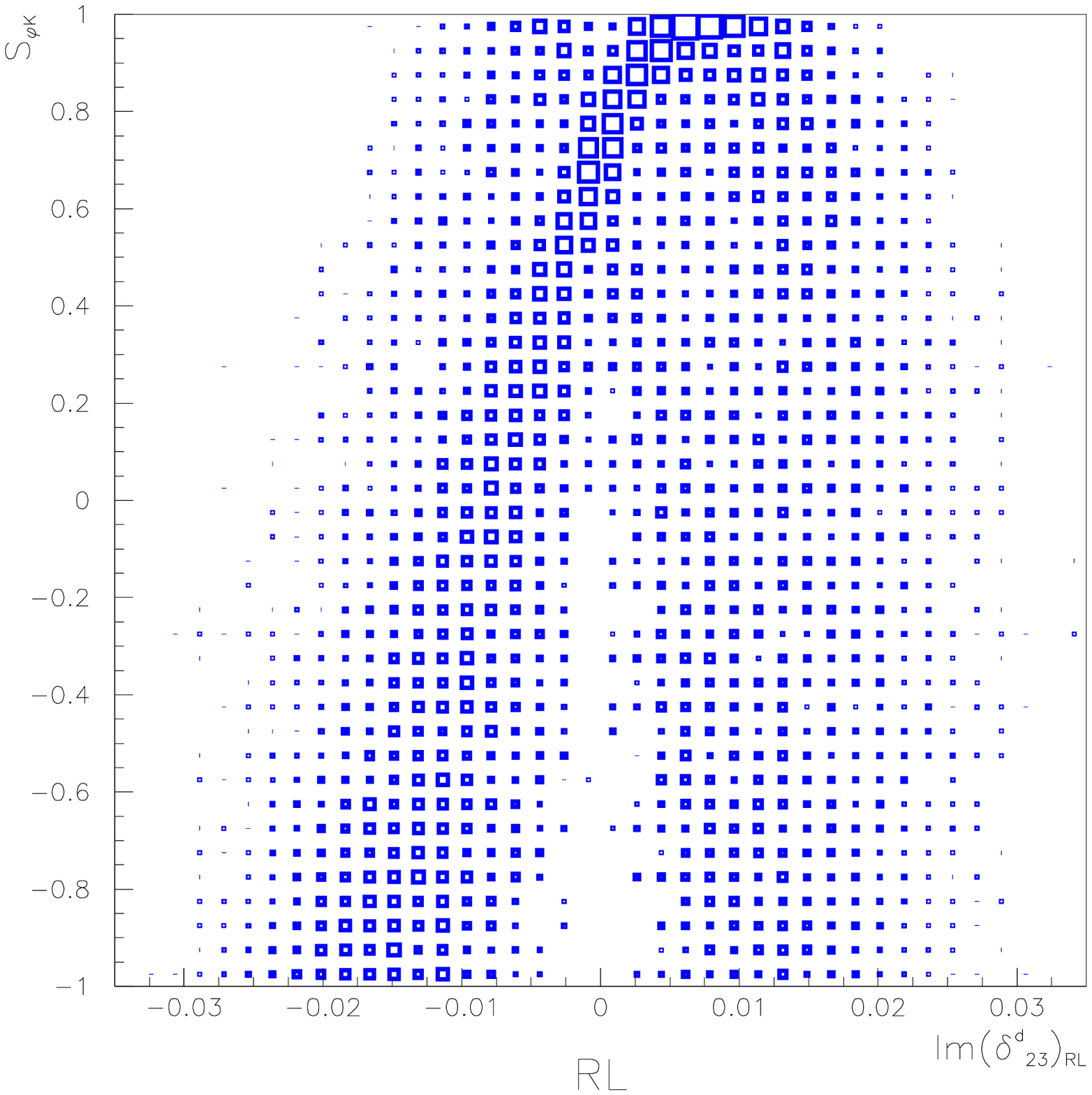} \\ 
    \end{tabular}
  \end{center}
  \caption{Correlations between $S_{\phi K}$ and Im$(\delta^d_{23})_{AB}$ 
    for  $AB=(LL,RR,LR,RL)$.
    }
  \label{fig:sinim1}
\end{figure}

\begin{figure}[t]
  \begin{center}
    \begin{tabular}{c c c}
      \includegraphics[width=0.31\textwidth]{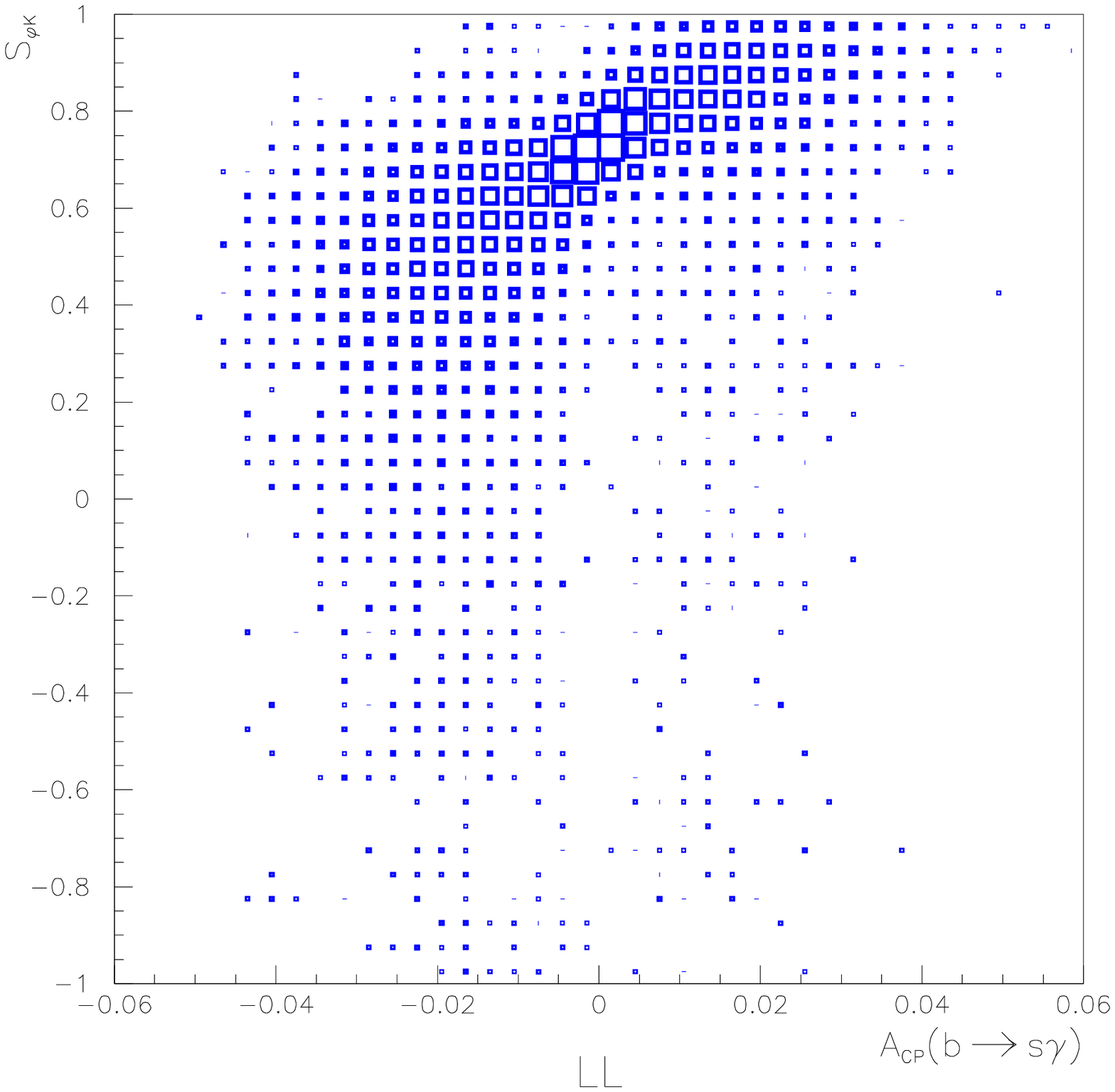} &
      \includegraphics[width=0.31\textwidth]{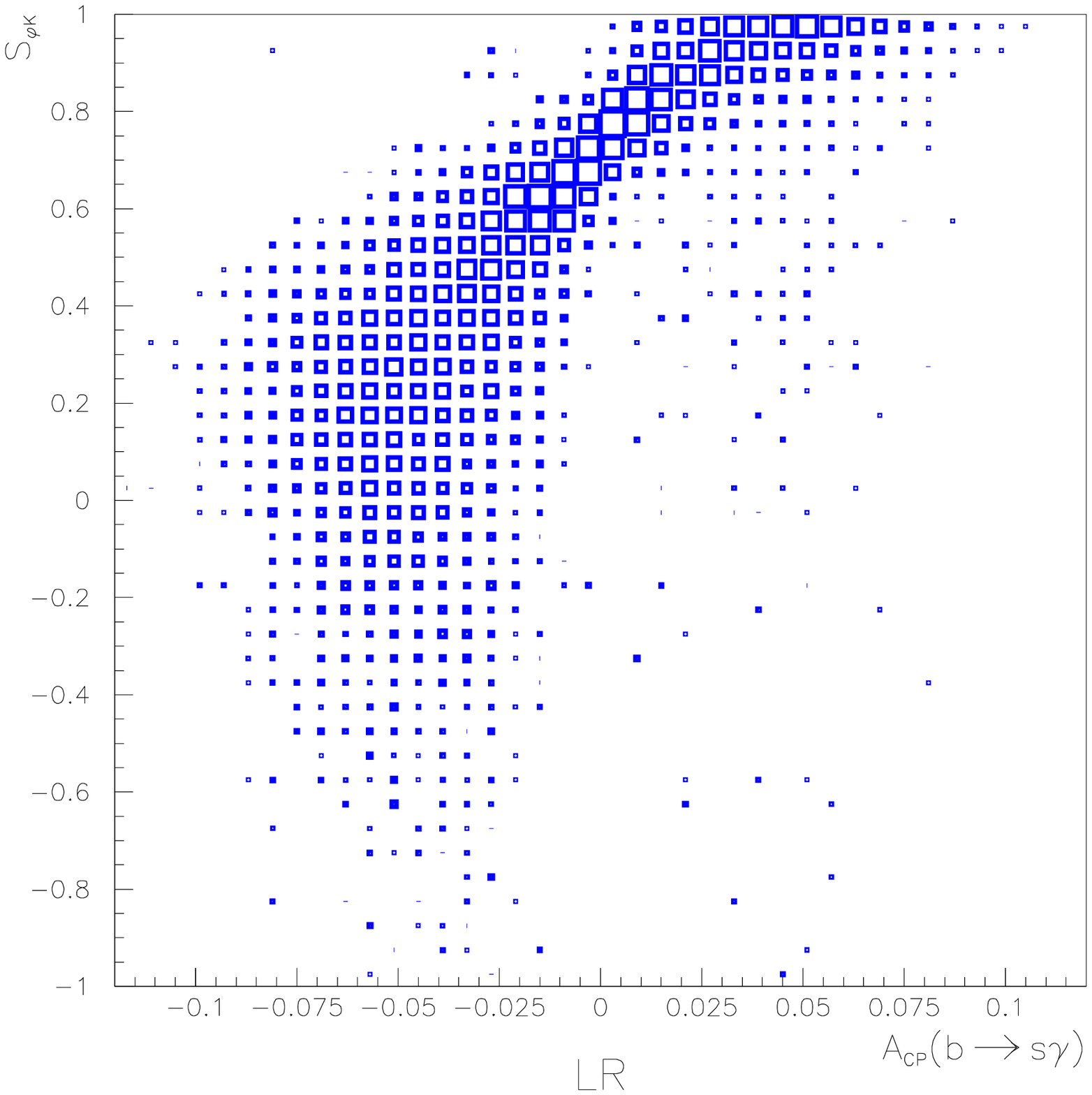} &
      \includegraphics[width=0.31\textwidth]{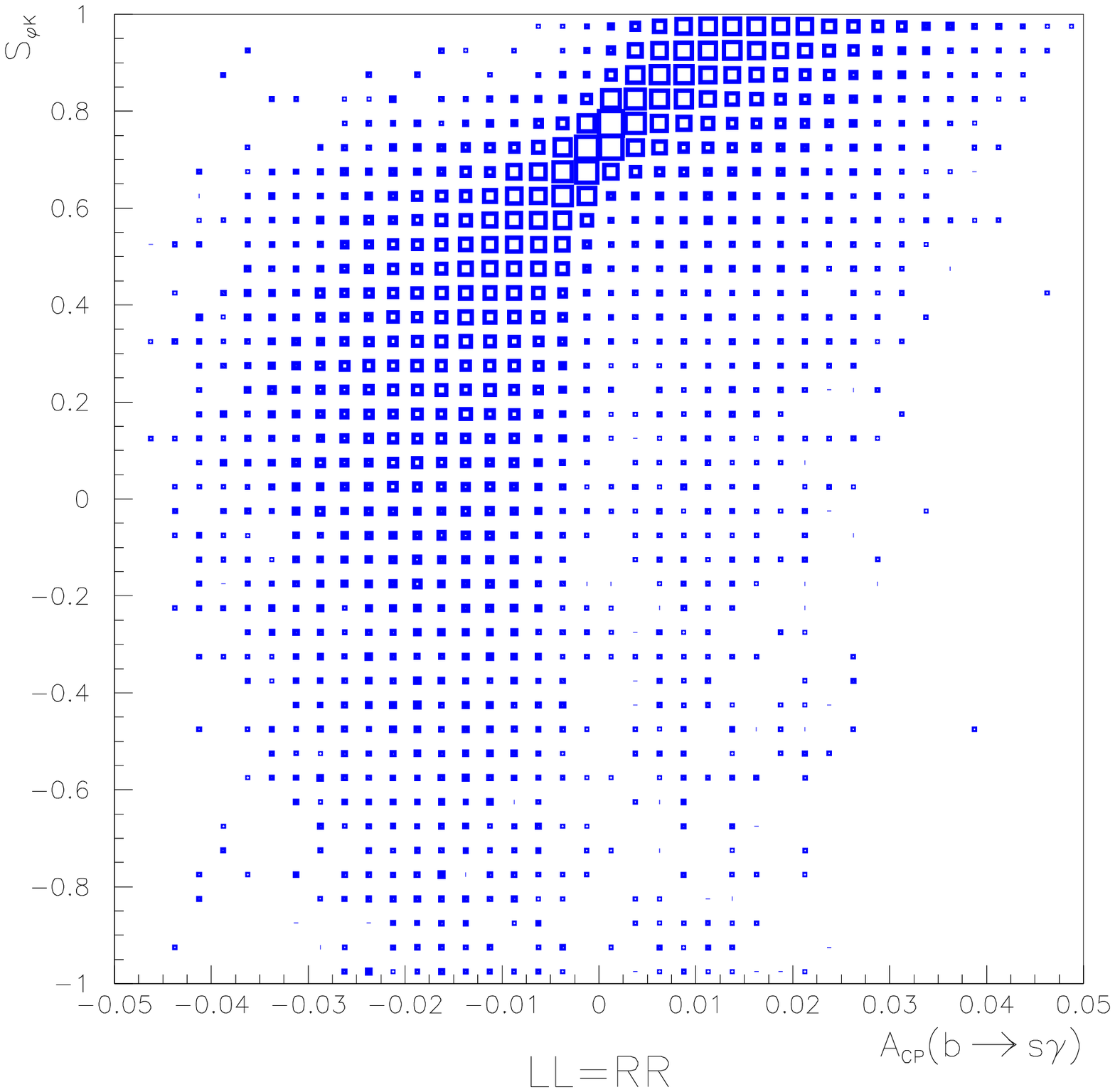}
      \\
    \end{tabular}
  \end{center}
  \caption{Correlation between $S_{\phi K}$ and $A_{CP}(b\to s\gamma)$
    for SUSY mass insertions $(\delta^d_{23})_{AB}$ with
    $AB=(LL,LR,LLRR)$.}
  \label{fig:sinabsg}
\end{figure}

\begin{figure}[t]
  \begin{center}
    \begin{tabular}{c c}
      \includegraphics[width=0.4\textwidth]{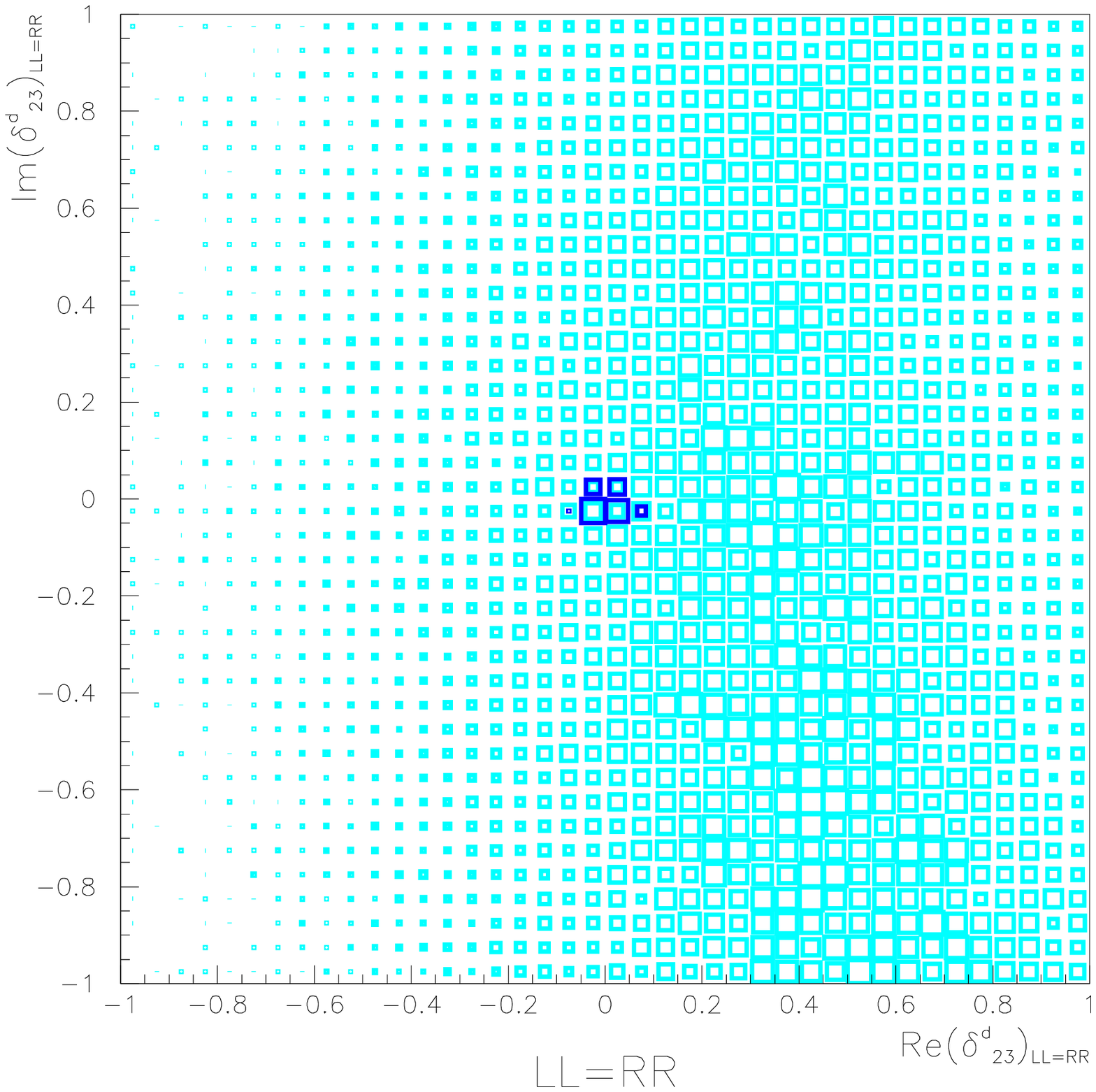} &
      \includegraphics[width=0.4\textwidth]{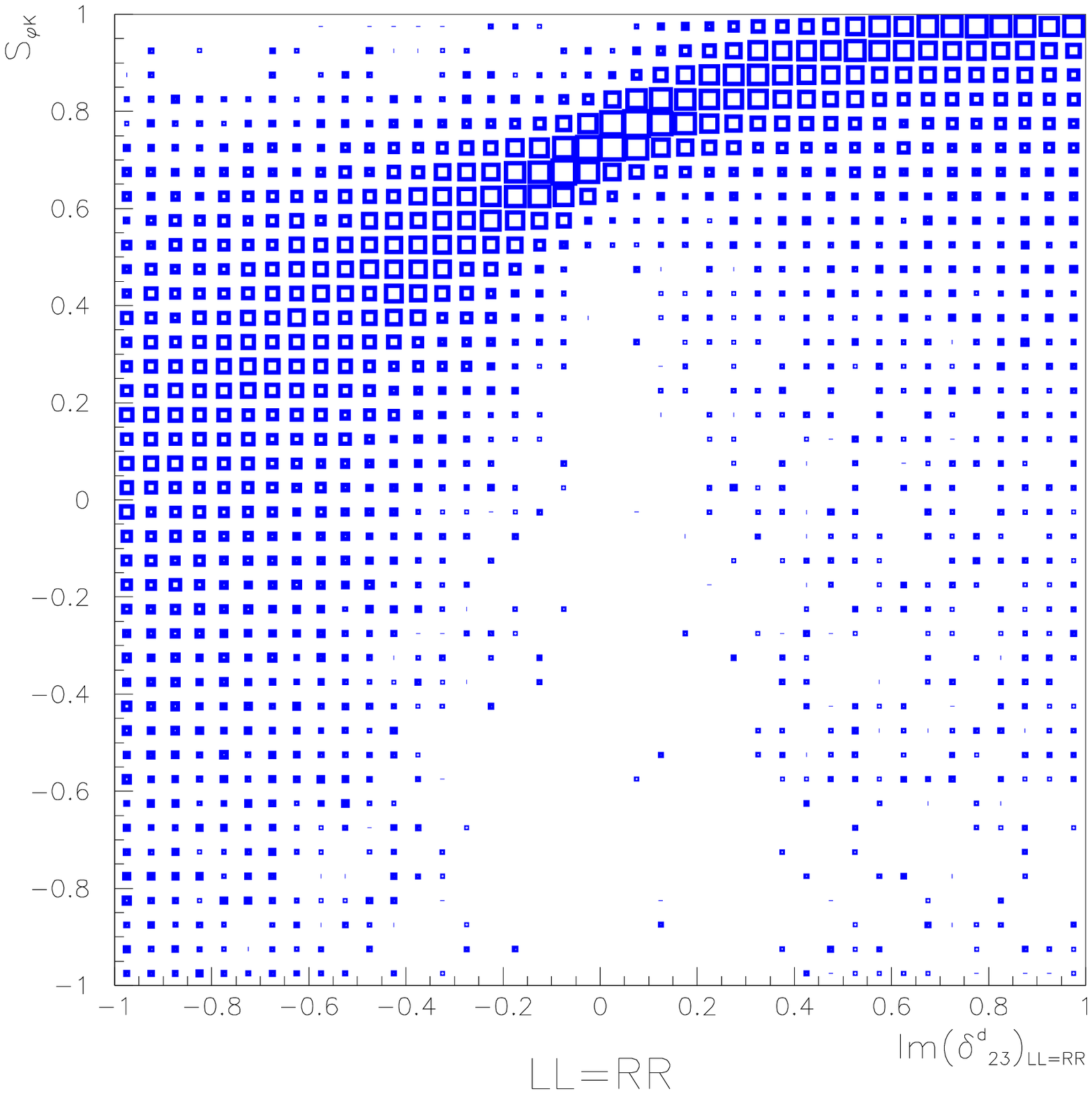}
      \\
    \end{tabular}
  \end{center}
  \caption{Same as in
figs.~\protect\ref{fig:ranges1}--\protect\ref{fig:sinim1} for the double 
insertion case $LL=RR$.}
  \label{fig:double}
\end{figure}

In figs.~\ref{fig:sinim1}--\ref{fig:sinabsg}, we study the
correlations of $S_{\phi K}$ with
Im$(\delta^d_{23})_{AB}$ and $A_{CP}(B\to X_s\gamma)$ for the various
SUSY insertions considered in the present analysis. The reader should
keep in mind that, in all the
results reported in figs.~\ref{fig:sinim1}--\ref{fig:sinabsg}, the
hadronic uncertainties affecting the estimate of $S_{\phi K}$ are not
completely under control. Low values of $S_{\phi K}$ can be more
easily obtained with helicity flipping insertions. A deviation from
the SM value for $S_{\phi K}$ requires a nonvanishing value of
Im$\,(\delta^d_{23})_{AB}$ (see figs.~\ref{fig:sinim1} and
\ref{fig:double}), generating, for those channels in which the SUSY
amplitude can interfere with the SM one, a $A_{CP}(B\to X_s\gamma)$ at
the level of a few percents in the LL and LL=RR cases, and up to the
experimental upper bound in the LR case (see fig.~\ref{fig:sinabsg})

Finally, fig.~\ref{fig:double} contains the same plots as
fig.~\ref{fig:ranges1}--\ref{fig:sinim1} 
in the case of the double mass insertion $(\delta^d_{23})_{LL}=(\delta^d_{23})_{RR}$.
In this case, the constraints are
still of order $1$, but the contribution to $\Delta M_s$ is huge, due
to the presence of operators with mixed chiralities. This can be seen
from the smallness of the dark region selected by imposing $\Delta
M_s<20$ ps$^{-1}$.

\section{Where to look for SUSY}
A crucial question naturally arises at this point: what are the more
promising processes to reveal some signal of low energy SUSY
among the FCNCs involving $b \to s$ transitions? 
For this purpose, it is useful to classify different ``classes of
MSSM'' according to the ``helicities'' $LL$, $RR$, etc, of the different $\delta_{23}^d$'s.

The BaBar and BELLE Collaborations have recently reported the
time-dependent CP asymmetry in $B_d(\bar B_d) \to \phi K_s$.
While $\sin 2 \beta$ as measured
in the $B \to J/\psi K_s$ channel is $0.734 \pm 0.054$ (in agreement
with the SM prediction~\cite{Stocchi:2002yi}), the combined result
from both collaborations for the corresponding $S_{\phi K}$ of $B_d
\to \phi K_s$ is $-0.39\pm 0.41$ \cite{Aubert:2002nx} with
a $2.7 \sigma$ discrepancy between the two results. In the SM, they should
be the same up to doubly Cabibbo suppressed terms. Obviously, one should be very cautious before accepting such result as a genuine indication of
NP.  Nonetheless, the negative value of $S_{\phi K}$ could be due to large
SUSY CP violating contributions. Then, one can wonder which $\delta$'s
are relevant to produce such enhancement and, even more important,
which other significant deviations from the SM could be detected.

\subsection{$RR$ and $LL$ cases}
\begin{figure}[t]
  \begin{center}
    \begin{tabular}{c c}
      \includegraphics[width=0.4\textwidth]{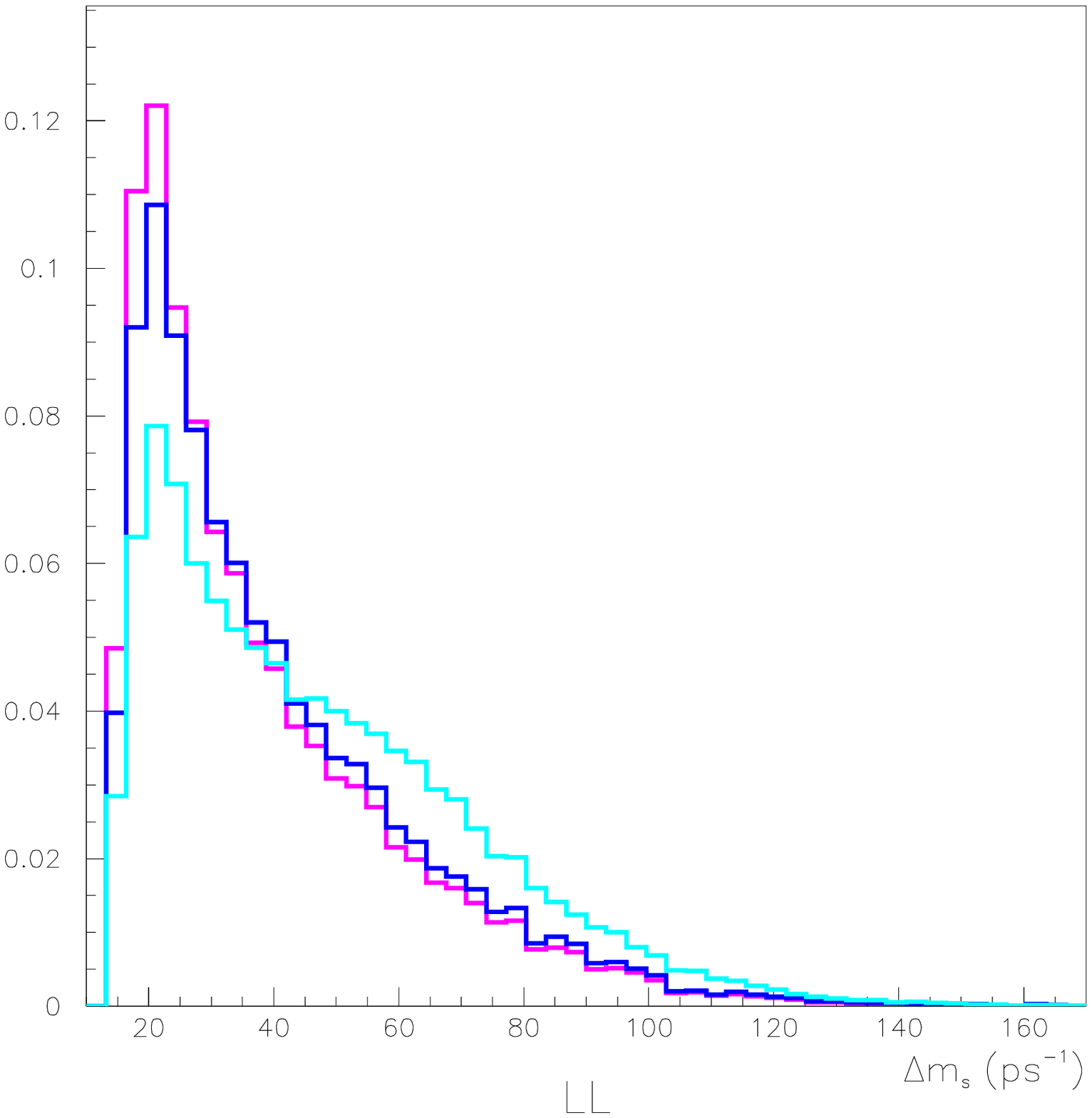} &
      \includegraphics[width=0.4\textwidth]{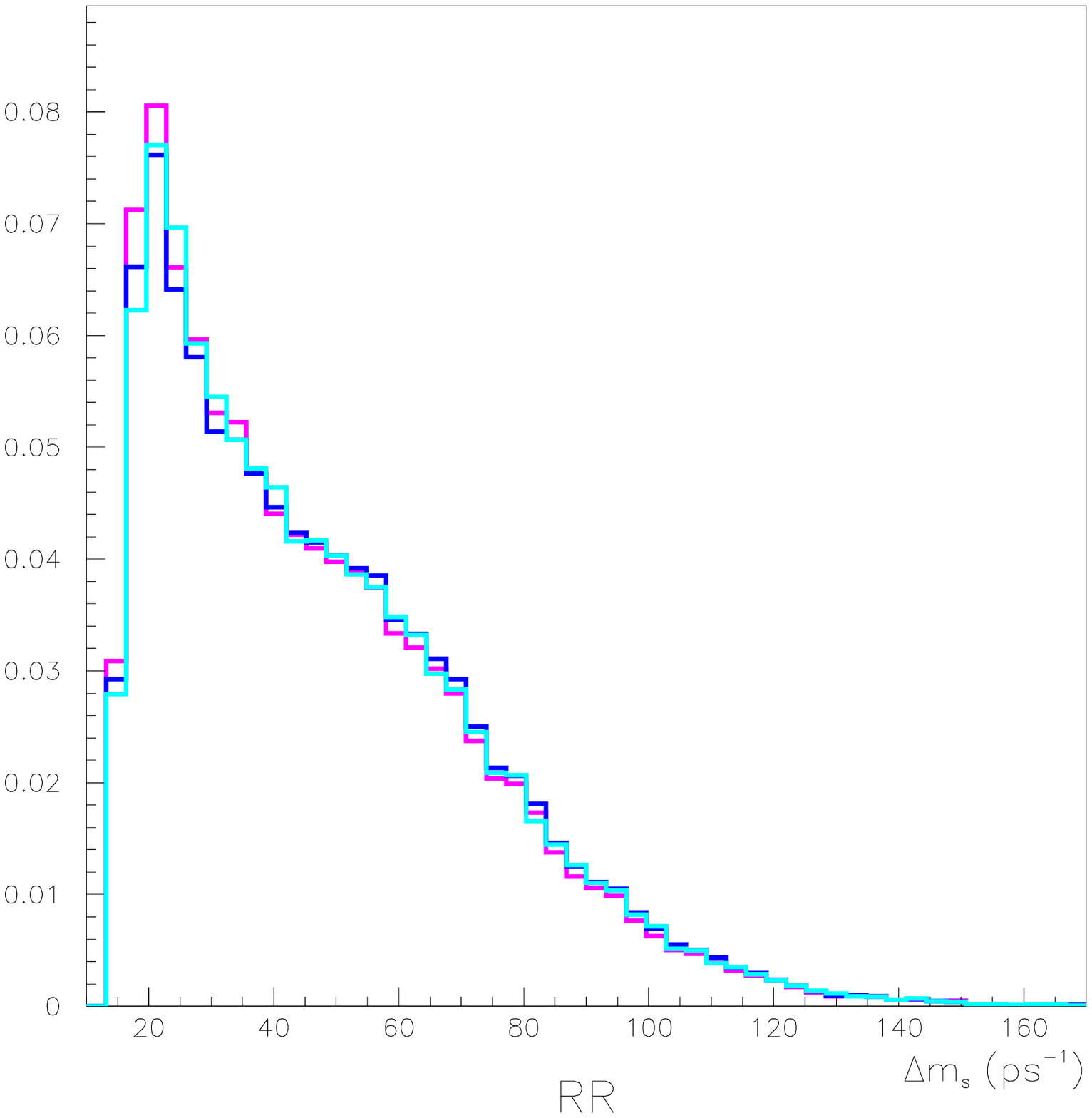} \\
      \includegraphics[width=0.4\textwidth]{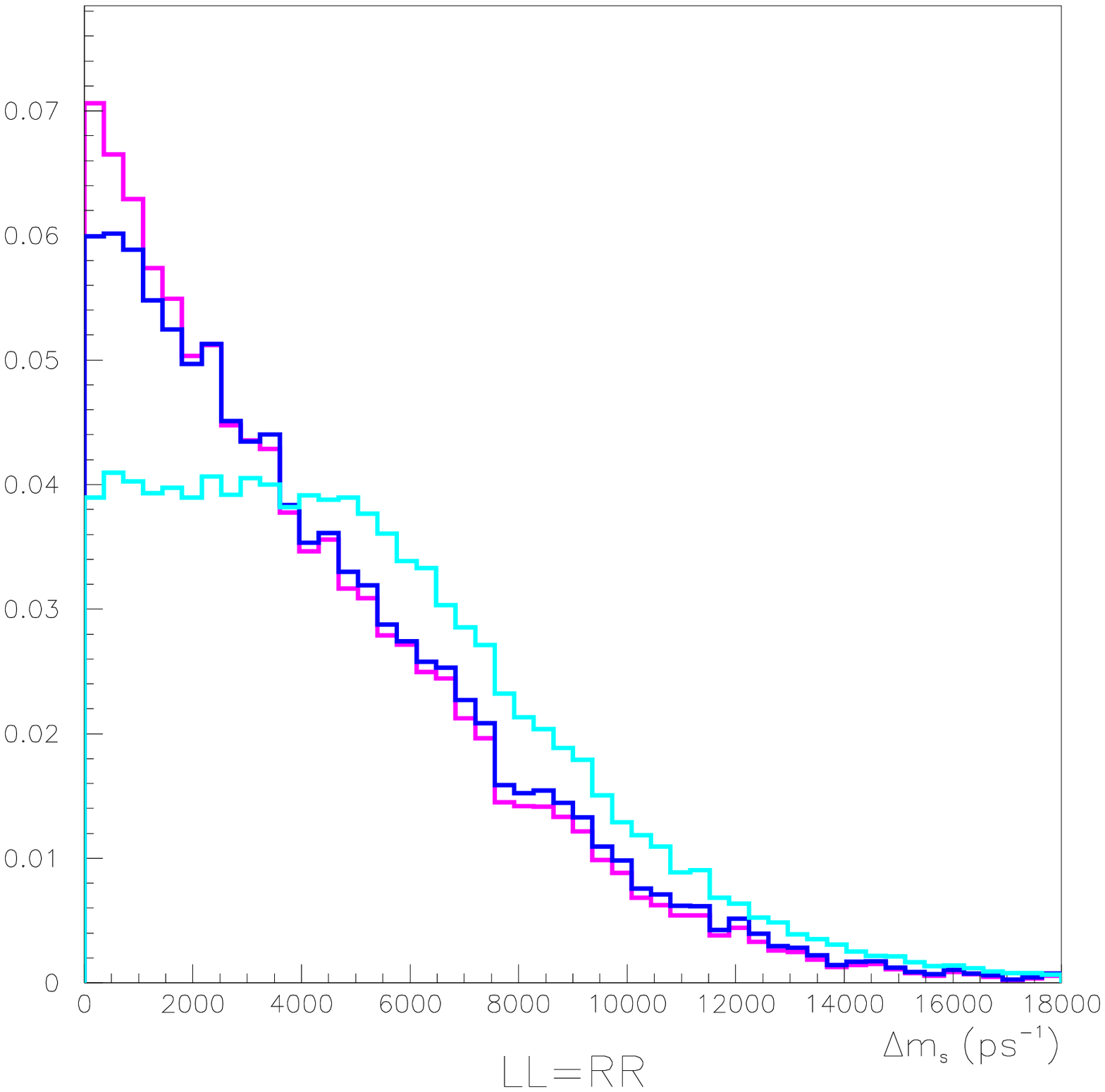} &
      \includegraphics[width=0.4\textwidth]{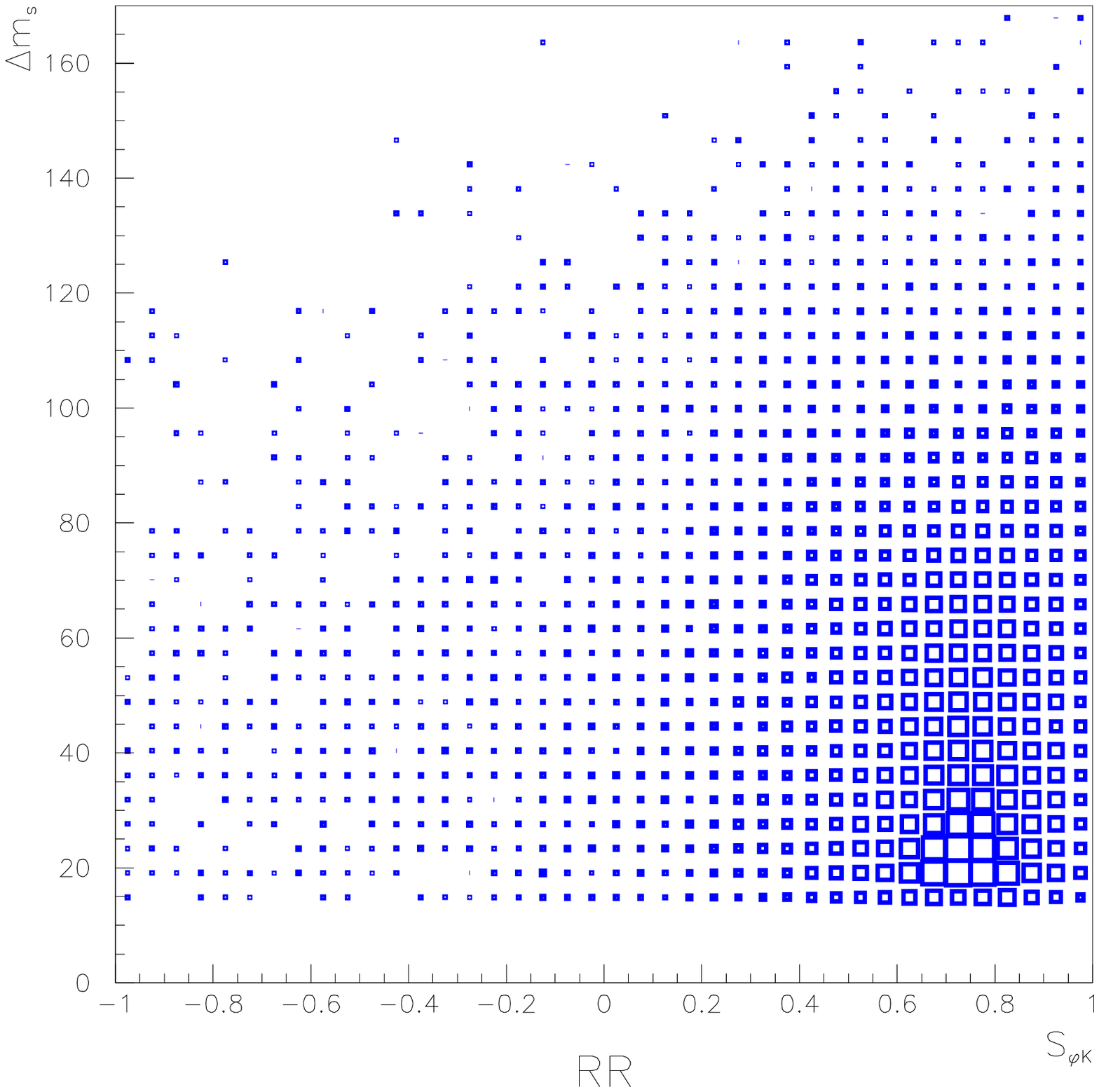} \\
      \\
    \end{tabular}
  \end{center}
  \caption{Distributions of $\Delta M_s$ for SUSY mass 
    insertions $(\delta^d_{23})_{AB}$ with $AB=(LL,RR,LLRR)$.
    Different curves correspond to the inclusion of constraints from
    $B\to X_s\gamma$ only (magenta), $B\to X_s l^+l^-$ only (cyan) and
    all together (blue).  Lower right: correlation between $\Delta
    M_s$ and $S_{\phi K}$ in the $RR$ case.  }
  \label{fig:dms}
\end{figure}

We start discussing the $RR$ case.
As shown in Fig.~\ref{fig:sinim1} (upper right), although values of
$S_{\phi K}$ in the range predicted by the SM are largely favoured,
still pure $\delta_{RR}$ insertions are able to give rise to a
negative $S_{\phi K}$ in agreement with the results of BaBar and BELLE
quoted above.  As for the $B_s - \bar B_s$ mixing,
the distribution of $\Delta M_s$ is peaked at the SM value, but it has a
long tail at larger values, up to $200$ ps$^{-1}$ for our choice of
the range of $\delta_{RR}$. In addition, we find that the expected
correlation requiring large $\Delta M_s$ for negative $S_{\phi K}$
is totally wiped out
by the large uncertainties (see fig.~\ref{fig:dms}, lower right). 
Hence, in the $RR$
case it is possible to have a strong discrepancy between $\sin
2\beta$ and $S_{\phi K}$ whilst $B_s-\bar B_s$ oscillations proceed as
expected in the SM (thus, being observable in the Run II of Tevatron).

To conclude the discussion of the $RR$ case, we expect the CP
asymmetry in $B \to X_s \gamma$ to be as small as in the
SM, while, differently from the SM, the
time-dependent CP asymmetry in the decay channel $B_s \to J/\psi \phi$
is expected to be large.

We now move on to discuss the $LL$ insertion.
A major difference with the previous case concerns the
SUSY contributions to $B \to X_s \gamma$. The $LL$ insertion 
contributes to the same operator which is responsible
for $B \to X_s \gamma$ in the SM and hence the SM and SUSY
amplitudes interfere. As a consequence, the rate tends to be larger
than the $RR$ case and, moreover, a CP asymmetry can be generated up
to $5 \%$ (see fig.~\ref{fig:sinabsg}, left). However, given the
uncertainties, the correlation of $A_{CP}(B\to X_s\gamma)$ with
$S_{\phi K}$ is not very stringent. As can be seen from the figure,
negative values of $S_{\phi K}$ do not necessarily correspond to
non-vanishing $A_{CP}(B\to X_s\gamma)$, although typical values are
around $2\%$. Also, the constraint coming from the present measurement
of the CP asymmetry is not very effective, as can be seen for instance
from the distribution of $\Delta M_s$ in fig.~\ref{fig:dms} which is
quite similar to the $RR$ case. Finally, one expects also in this case
to observe CP violation in $B_s \to J/\psi \phi$ at hadron colliders.

\subsection{$LR$ and $RL$ cases}
In these cases negative values of $S_{\phi K}$ can be easily obtained
(although a positive $S_{\phi K}$ is slightly favoured,
cfr.~Fig.~\ref{fig:sinim1}, bottom row). The severe bound on the $LR$
mass insertion imposed by BR$(B \to X_s \gamma)$ (and $A_{CP}(B \to
X_s \gamma)$ in the $LR$ case) prevents any enhancement of the $B_s -
\bar B_s$ mixing as well as any sizeable contribution to $A_{CP}(B_s
\to J/\psi \phi)$. On the other hand, $A_{CP}(B \to X_s \gamma)$ as
large as $5$--$10$ \% is now attainable (Fig.~\ref{fig:sinabsg}, upper
right), offering a potentially interesting hint for NP.

Notice that the $LR$ mass insertion contributes to $b_R \to s_L
\gamma$, much like the SM. The interference with the SM amplitude produces the 'hole' in fig.~\ref{fig:ranges1}, lower left.
On the contrary, the $RL$ mass insertion contributes to $b_L \to
s_R \gamma$ and thus it does not interfere.
Consequently, the CP asymmetry is as small as in the SM and the $RL$ mass
insertion is less constrained than the $LR$ one by $B \to X_s \gamma$,
allowing for negative values of $S_{\phi K}$ to be produced more
easily.

\subsection{Double mass insertion: $(\delta_{23})_{LL}
  =(\delta_{23})_{RR}$ case}
The main feature of this case is the huge enhancement of $\Delta M_s$
which is made possible by the contribution of the double insertion $LL$
and $RR$ in the box diagrams to operators with mixed chiralities
(Fig.~\ref{fig:dms}, lower left). Differently from all the previous
cases, we are facing a situation here where $A_{CP}(B \to \phi K_s)$ at
its present experimental value should be accounted for by the presence
of SUSY, while $\Delta M_s$ could be so large that the $B_s - \bar
B_s$ mixing could escape detection not only at Tevatron, but even at
BTeV or LHCB. Hence, this would be a case for remarkable signatures of
SUSY in $b \to s$ physics.

Finally, we remark that in the $LR$ and $RL$
cases, since for $m_{\tilde g}=m_{\tilde q}=350$ GeV the constraints
on the $\delta_{23}^d$'s are of order $10^{-2}$, the same phenomenology in $\Delta B=1$ processes can be obtained
at larger values of mass insertions and of squark and gluino masses,
while contributions to $\Delta B=2$ processes become more
important for larger masses. In the remaining cases, where the limits
on $\delta_{23}^d$ at $m_{\tilde g}=m_{\tilde q}=350$ GeV are of order
$1$, the SUSY effects clearly weaken when going to higher values of
sparticle masses. 

\section{Outlook}
 
Our results confirm that FCNC and CP violation in
physics involving $b \to s$ transitions still offer
opportunities to disentangle effects genuinely due to NP. In
particular the discrepancy between the amounts of CP violation in the
two $B_d$ decay channels $J/\psi K_s$ and $\phi K_s$ can be accounted
for in the MSSM while respecting all the existing constraints in $B$
physics, first of all the $BR(B \to X_s \gamma)$. The relevant question
is then which processes offer the best chances to provide other hints
of the presence of low-energy SUSY.

First,  it is mandatory to further assess the
time-dependent CP asymmetry in the decay channel $B \to \phi K_s$.
If the measurement will be
confirmed, then this process would become decisive in discriminating
among different MSSM realizations. Although, as we have seen, it is
possible to reproduce the negative $S_{\phi K}$ in a variety of different
options for the SUSY soft breaking down squark masses, the allowed
regions in the SUSY parameter space are more or
less tightly constrained according to the kind of $\delta_{23}^d$ mass
insertion which dominates.

In order of importance, it then comes the measurement of the $B_s - \bar B_s$ mixing. Finding $\Delta M_s$ larger than $20$ ps$^{-1}$ would hint at
NP. $RR$ or $LL$ could account for a $\Delta M_s$
up to $200$ ps$^{-1}$. Larger values would call for the double
insertion $LL=RR$ to ensure such a huge enhancement of $\Delta M_s$
while respecting the constraint on $BR(B \to X_s \gamma)$. An
interesting alternative would arise if $\Delta M_s$ is
found as expected in the SM while, at the same time, $S_{\phi K}$ is
confirmed to be negative. This scenario would favour the $LR$
possibility, even though all other cases but $LL=RR$ do not
necessarily lead to large $\Delta M_s$.

Keeping to $B_d$ physics, we point out
that the CP asymmetry in $B \to X_s \gamma$ remains of utmost
interest.  This asymmetry is so small in the SM that it
should not be possible to detect it. We have seen that in particular
with $LR$ insertions such asymmetry can be enhanced up to
$10$ \% making it possibly detectable in a not too distant future.

Finally, once we will have at disposal large amounts of $B_s$,
it will be of great interest to study processes which
are mostly CP conserving in the SM, while possibly receiving large contributions from SUSY. In the SM the amplitude for $B_s - \bar
B_s$ mixing does not have an imaginary part up to doubly Cabibbo suppressed terms and decays like $B_s \to J/\psi \phi$ also have a negligible amount of CP violation.
Quite on the contrary, if the measured negative $S_{\phi K}$ is due to
a large, complex $\delta_{23}^d$ mass insertion, we expect some of the
above processes to exhibit a significant amount of CP violation. In
particular, in the case of $RR$ insertions, both the $b \to s$
amplitudes and the $B_s$ mixing would receive non negligible
contributions from Im$\,\delta_{23}^d$, while, if the $LR$ insertions is
dominant, we do not expect any sizable contribution to $B_s$ mixing. Still, the SUSY contribution to CP violation in the $B_s \to J/\psi \phi$
decay amplitude could be fairly large.


\section*{References}
\begin{thebibliography}{99}
\bibitem{antichibagger}
F.~Gabbiani {\it et al.},
Nucl.\ Phys.\ B {\bf 477} (1996) 321
[arXiv:hep-ph/9604387].

\bibitem{Ciuchini:2002uv}
M.~Ciuchini {\it et al.},
Phys.\ Rev.\ D {\bf 67} (2003) 075016
[arXiv:hep-ph/0212397] (n.b.: versions lower than 3 are incorrect).

 \bibitem{hall}
L.~J.~Hall, V.~A.~Kostelecky and S.~Raby,
Nucl.\ Phys.\ B {\bf 267} (1986) 415.

\bibitem{Casas:1996de}
J.~A.~Casas and S.~Dimopoulos,
Phys.\ Lett.\ B {\bf 387} (1996) 107
[arXiv:hep-ph/9606237].

\bibitem{Becirevic:2001jj}
D.~Becirevic {\it et al.},
Nucl.\ Phys.\ B {\bf 634} (2002) 105
[arXiv:hep-ph/0112303].

\bibitem{Ciuchini:1998ix}
M.~Ciuchini {\it et al.},
JHEP {\bf 9810} (1998) 008
[arXiv:hep-ph/9808328].

\bibitem{Aubert:2002nx}
B.~Aubert {\it et al.}  [BABAR Collaboration],
arXiv:hep-ex/0207070.

\bibitem{Stocchi:2002yi}
A.~Stocchi,
arXiv:hep-ph/0211245.

\bibitem{Gambino:2001ew}
P.~Gambino and M.~Misiak,
Nucl.\ Phys.\ B {\bf 611} (2001) 338
[arXiv:hep-ph/0104034].

\bibitem{BBNS}
M.~Beneke, G.~Buchalla, M.~Neubert and C.~T.~Sachrajda,
Phys.\ Rev.\ Lett.\  {\bf 83} (1999) 1914
[arXiv:hep-ph/9905312];
Nucl.\ Phys.\ B {\bf 606} (2001) 245
[arXiv:hep-ph/0104110].

\bibitem{Charming}
M.~Ciuchini {\it et al.},
Phys.\ Lett.\ B {\bf 515} (2001) 33
[arXiv:hep-ph/0104126];
arXiv:hep-ph/0208048.

\bibitem{Ciuchini:1997hb}
M.~Ciuchini {\it et al.},
Nucl.\ Phys.\ B {\bf 501} (1997) 271
[arXiv:hep-ph/9703353].

\bibitem{c8g}
A.~L.~Kagan,
Phys.\ Rev.\ D {\bf 51} (1995) 6196
[arXiv:hep-ph/9409215];\\
M.~Ciuchini, E.~Gabrielli and G.~F.~Giudice,
Phys.\ Lett.\ B {\bf 388} (1996) 353
[Erratum-ibid.\ B {\bf 393} (1997) 489]
[arXiv:hep-ph/9604438].

\bibitem{Ciuchini:2000de}
M.~Ciuchini {\it et al.},
JHEP {\bf 0107} (2001) 013
[arXiv:hep-ph/0012308].

\end{thebibliography}
\end{document}